\documentclass[10pt,conference,letterpaper]{IEEEtran}
\IEEEoverridecommandlockouts
\usepackage[T1]{fontenc}
\usepackage{cite}
\usepackage{amsmath,amssymb,amsfonts}
\usepackage{algorithm}
\usepackage[noend]{algpseudocode}
\usepackage{booktabs}
\usepackage{multirow}
\usepackage[svgnames, table]{xcolor}
\usepackage{tabulary}
\usepackage{caption}
\usepackage{underscore}
\usepackage{quantikz}
\usepackage{mathtools}
\usepackage{graphicx}
\usepackage{textcomp}
\usepackage{dcolumn} 
\usepackage{bm} 
\usepackage{braket}
\usepackage[normalem]{ulem}
\usepackage[]{hyperref} 
\hypersetup{
    colorlinks = true,
    citecolor = magenta,
    linkcolor = purple
    }
\usepackage[capitalise]{cleveref}
\crefname{figure}{Fig.}{Figs.}
\crefname{equation}{}{}
\Crefname{equation}{Equation}{Equations}

\def\BibTeX{{\rm B\kern-.05em{\sc i\kern-.025em b}\kern-.08em
    T\kern-.1667em\lower.7ex\hbox{E}\kern-.125emX}}

\begin{document}
\bstctlcite{IEEEexample:BSTcontrol}

\title{
Space-Time Tradeoffs of Pauli-Based Computation in Distributed qLDPC Architectures
\thanks{
This work was supported by the JST Moonshot R\&D program under Grant Numbers JPMJMS2061, JPMJMS256E, JPMJMS226C, and JPMJMS256K.
}
}

\author{
\IEEEauthorblockN{
Naphan Benchasattabuse\IEEEauthorrefmark{1},
Michal Hajdu\v{s}ek\IEEEauthorrefmark{1},
and Rodney Van Meter\IEEEauthorrefmark{2}}\\

\IEEEauthorblockA{\IEEEauthorrefmark{1}\textit{Graduate School of Media Design, Keio University, Kanagawa, Japan}}
\IEEEauthorblockA{\IEEEauthorrefmark{2}\textit{Faculty of Environment and Information Studies, Keio University, Kanagawa, Japan}\\
\{whit3z,michal,rdv\}@sfc.wide.ad.jp}
}

\thispagestyle{plain}
\pagestyle{plain}

\maketitle

\begin{abstract}
Pauli-based computation (PBC) provides a universal framework for executing fault-tolerant quantum algorithms using Pauli measurements and magic states.
In monolithic architectures, the serialized nature of PBC directly ties runtime to a circuit's T-gate count, making it slow on metrics like circuit depth.
However, in distributed quantum computing (DQC), the primary bottleneck is remote Bell pair generation.
We investigate the tradeoff between error-correcting code block size and execution time of PBC within the Q-Fly architecture at intermediate scale, limiting individual node capacities to reflect near-term constraints while supplying abundant network nodes to minimize routing and compilation effects.
We find that large qLDPC code blocks outperform the surface code baseline in terms of execution time by up to an order of magnitude when evaluated against quantum optimization algorithms.
By moving groups of qubits to free nodes to bypass the sequential bottleneck of PBC, the large-block architecture minimizes network operations and achieves faster overall execution.
This demonstrates that PBC is a competitive model in the distributed regime, establishing it as a practical compilation baseline for qLDPC systems before invoking more efficient transversal or homological gates.
\end{abstract}


\section{Introduction}
\label{sec:introduction}

Space-time tradeoffs of computation have been fundamental since the early days of computer science~\cite{bennettTimeSpaceTradeOffs1989,hopcroftTimeSpace1977,borodinTimeSpaceTradeoffSorting1982}.
In the quantum domain, this same tradeoff emerges naturally in the circuit model between circuit depth and ancillary qubit space.
Substantial prior work explores this relationship under various computational models, spanning from shallow-depth circuit complexity classes~\cite{hoyerQuantumFanoutPowerful2005,mooreQuantumCircuitsFanout1999,grierQuantumThresholdPowerful2024,greenCountingFanoutComplexity2002,miyazakiAnalysisTradeoffSpatial2015} to measurement-based computation.
It is especially prominent in Pauli-based and fault-tolerant quantum computation (FTQC)~\cite{fowlerTimeoptimalQuantumComputation2013,litinskiGameSurfaceCodes2019,raussendorfTopologicalFaulttoleranceCluster2007}, where one can directly trade space for time through topological deformations.
Because distributed quantum computation (DQC) aims to scale up FTQC by networking multiple discrete nodes~\cite{buhrmanDistributedQuantumComputing2003,ciracDistributedQuantumComputation1999,nickersonTopologicalQuantumComputing2013,monroeLargescaleModularQuantumcomputer2014,jnaneMulticoreQuantumComputing2022,angARQUINArchitecturesMultinode2024}, deliberately trading space for time in these systems becomes the key strategy to accelerate our path toward the quantum advantage regime.

The key enabler of DQC is the ability to generate shared entangled states across networked quantum processing units (QPUs)~\cite{wehnerQuantumInternetVision2018,awschalomDevelopmentQuantumInterconnects2021}.
These QPUs communicate via interconnects, and the speed of generating this inter-node entanglement dictates the primary performance bottleneck of the entire distributed architecture.
While local operations are often orders of magnitude faster than these interconnects, this communication delay represents merely a constant slowdown in the asymptotic limit.
Consequently, mitigating this delay is of little interest from a circuit complexity perspective.
However, in practical settings, this constant factor governs the true runtime, making it a critical hurdle that we must explicitly address.

Most works addressing DQC compilation over slow interconnects assume a static input circuit comprising arbitrary single-qubit gates and CNOTs~\cite{zhangSwitchQNetOptimizingDistributed2025,burtMultilevelFrameworkPartitioning2026,chenCircuitPartitioningTransmission2025,russoTeleSABRELayoutSynthesis2025,escofetRevisitingMappingQuantum2025,maoQubitAllocationDistributed2023,sundaramDQCQRDistributingRouting2025,andres-martinezDistributingCircuitsHeterogeneous2024,kaurOptimizedQuantumCircuit2025,bandiniOptimizedCompilationDistributed2026}.
The two main approaches to tackle this are assigning qubits to specific QPUs via graph partitioning to minimize cross-node CNOT operations, or exploring qubit routing via telegates and teledata.
Within these routing methods, telegates generally equate a single Bell pair to a single CNOT, while teledata physically teleports the data qubit across nodes to interact with multiple targets.
Although a single Bell pair can natively implement remote multi-target CNOT gates (fan-out) without moving the data qubit~\cite{eisertOptimalLocalImplementation2000}, prior works exploring distributed fan-out often rely on multi-partite GHZ states instead~\cite{yimsiriwattanaGeneralizedGHZStates2004,llovoNetworkAssistedCollectiveOperations2025,lokeDistributedQuantumComputing2026}.
Fundamentally, these methods merely partition or route pre-defined circuits, offering no insight into how to design better circuits from the ground up.
Consequently, treating compilation primarily as a static mapping obscures the true cost of DQC, which is the overall consumption of entangled states.
Additionally, these models often ignore the operational realities of FTQC, failing to account for  the seralizaed nature of Pauli-based computation (PBC)~\cite{bravyiTradingClassicalQuantum2016} and the significant overhead of non-Cliffordness~\cite{litinskiMagicStateDistillation2019}.
Efforts to explicitly minimize this entanglement cost are typically isolated to network contexts or the measurement-based quantum computation (MBQC) model for graph state preparation~\cite{miccicheQuantumHamletsDistributed2026,senMultipartiteEntanglementDistribution2025,sharmaMinimisingNumberEdges2026}.
The main question thus remains: how do we design circuits that achieve an optimal space-time balance for DQC while properly accounting for the slow nature of interconnects and the requirements of fault-tolerant execution?

In this work, we present techniques to design and construct circuits tailored for DQC.
Because practical FTQC demands measurement-based execution, we explicitly target architectures employing high-rate quantum codes operated via PBC~\cite{yoderTourGrossModular2025,heExtractorsQLDPCArchitectures2025,websterPinnacleArchitectureReducing2026,mundadaHeterogeneousArchitecturesEnable2026,cainShorsAlgorithmPossible2026}.
In particular, we utilize qLDPC codes equipped with extractors---a generic surgery approach with a fixed ancilla system proposed by He et al.~\cite{heExtractorsQLDPCArchitectures2025}.
While high-rate codes drastically reduce spatial overhead, they often suffer from slow sequential gate operations that scale with the code distance unless mitigated by code-specific crafted gates, which we avoid to maintain generality.
To evaluate our approach, we use the Q-Fly architecture~\cite{sakumaQFlyOpticalInterconnect2025}, a hierarchically interconnected system mimicking the dragonfly topology of modern supercomputers.
We establish a projected, realistic hardware baseline to isolate the performance of our compilation techniques.
Specifically, we assume each node has local access to magic states, limited communication buffers for Bell pairs, and a sufficient number of QPUs to prevent artificial bottlenecks.

Rather than targeting resource-heavy cryptographic benchmarks like Shor's algorithm, we focus on near-term FTQC optimization applications.
We analyze the quantum approximate optimization algorithm (QAOA)~\cite{farhiQuantumApproximateOptimization2014} and decoded quantum interferometry (DQI)~\cite{jordanOptimizationDecodedQuantum2025}.
These algorithms provide a rich set of common subroutines and have demonstrated quantum advantage over classical heuristics~\cite{boulebnaneSolvingBooleanSatisfiability2024,shaydulinEvidenceScalingAdvantage2024,khattarVerifiableQuantumAdvantage2025}.
By fixing the number of nodes, node groups, and physical qubits per node, we directly compare the runtime of our custom-designed circuits against an active volume surface code baseline constructed for state-of-the-art monolithic architecture~\cite{litinskiActiveVolumeArchitecture2022}.
Our approach demonstrates advantage in all test cases, sometimes leading to an order of magnitude saving in execution time, showcasing the substantial performance improvements that architecture-aware circuit design can unlock for distributed systems using high-rate codes.

We begin by introducing distributed quantum architectures and detailing the specific Q-Fly model (\cref{sec:dqc-architecture}).
We then formalize our evaluation framework by justifying specific parameter choices, defining hardware constraints, and discussing scalability (\cref{sec:arch-constraints}).
Next, we outline the target optimization algorithms, providing a brief background on the specific subroutines required for DQI and QAOA (\cref{sec:optimization-algs}).
We then detail space-time circuit optimization techniques explicitly tailored for the Q-Fly architecture within the PBC framework (\cref{sec:circuit-optimization-techniques}).
We perform a comprehensive resource estimation to establish an upper bound for our custom circuits and benchmark our approach against a baseline active volume surface code model (\cref{sec:resource-estimation}).
Finally, we discuss the broader implications of this work and highlight future directions for the co-design of architectures and circuits (\cref{sec:discussion}).

\section{Distributed quantum architecture}
\label{sec:dqc-architecture}

Distributed quantum computation (DQC) scales processing power by connecting multiple quantum computers capable of sharing generic entangled states, such as Bell or GHZ states, across nodes.
This work specifically targets modular quantum computers, also known as data-center scale multicomputers.
In this paradigm, nodes reside in close physical proximity, eliminating the need for complex quantum repeaters~\cite{azumaQuantumRepeatersQuantum2023}.
We assume photonic interconnects mediate the network to generate Bell pairs.
Typically, two nodes first generate memory-photon entanglement, and these photons are then routed to a Bell state analyzer (BSA).
The BSA performs a Bell state measurement, which is a probabilistic process that heralds entanglement between the stationary memory qubits.

A key design consideration for DQC is the network topology, which dictates how nodes connect and directly share Bell pairs.
Standard topologies include simple rings, where each node connects to its two immediate neighbors, or linear chains bounded by endpoints.
Other common configurations include grids, star graphs, and all-to-all connectivity.
When evaluating these topologies, one must consider the hardware overhead of the required BSAs.
Because BSAs are resource-intensive components, fully connected topologies or dense grids demand a prohibitively large number of them.
Reconfigurable switch architectures, inspired by classical supercomputer networks, offer a way to minimize this BSA count.
Prominent examples include Clos networks~\cite{shapourianQuantumDataCenter2025,pouryousefBenchmarkingQuantumData2026}, fat-trees~\cite{choiScalableQuantumNetworks2023}, and the dragonfly-inspired Q-Fly architecture~\cite{sakumaQFlyOpticalInterconnect2025}.
Because these architectures are hierarchical, photons must route through multiple switches, which inherently drops the entanglement rate due to cumulative coupling losses.
We select the Q-Fly topology because its symmetric, hierarchical design minimizes both the hop count and the required switch sizes while maintaining uniform bandwidth across all links, unlike heavily level-based structures such as fat-trees or Clos networks.

\begin{figure}[t]
    \centering
    \includegraphics[width=\columnwidth]{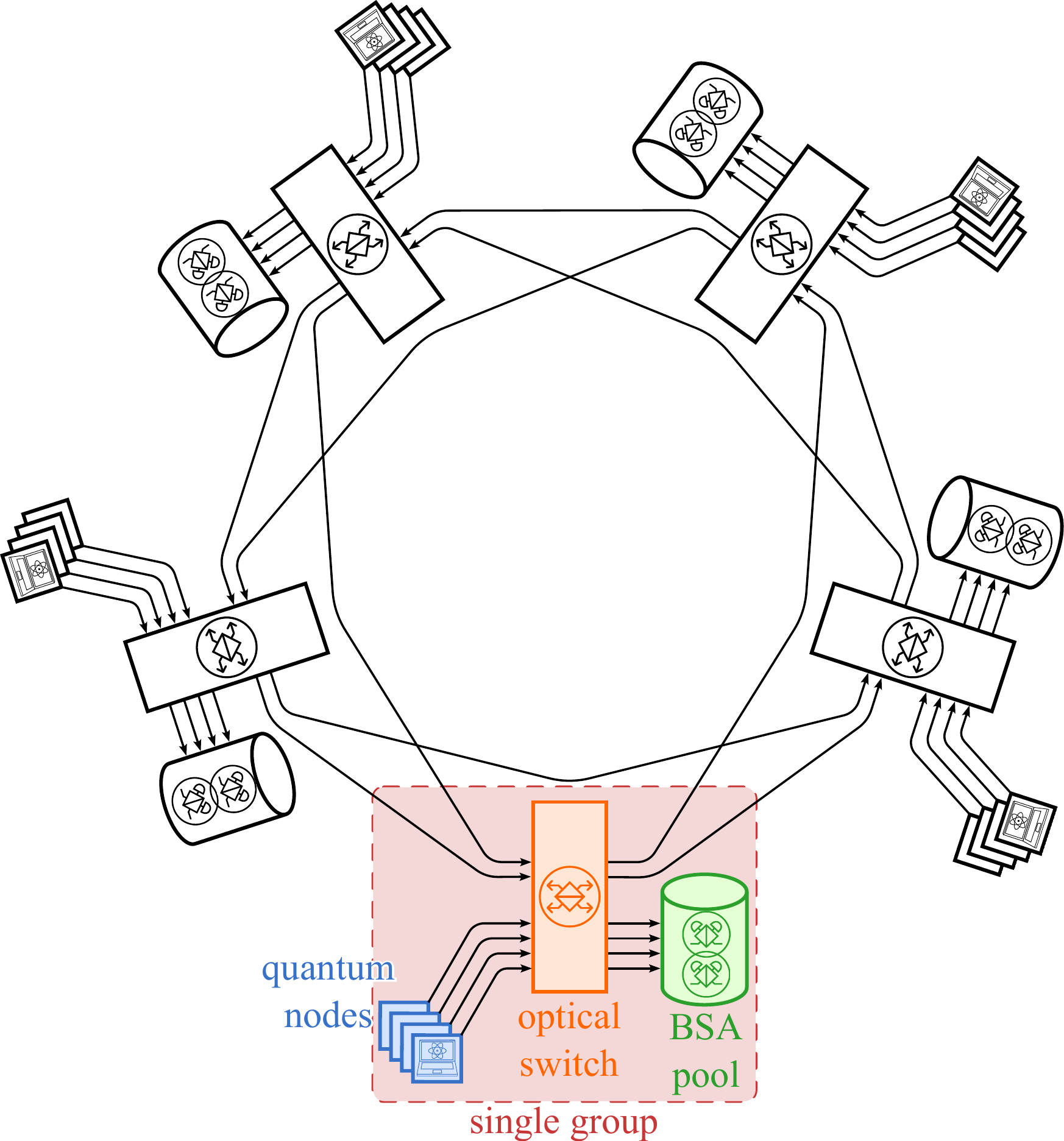}
    \caption{Example of a Q-Fly architecture with five groups, each composed of four computational quantum nodes, an optical switch and a pool of BSAs.}
    \label{fig:qFly}
\end{figure}

The Q-Fly architecture, illustrated in \cref{fig:qFly}, operates on a two-level hierarchy.
At the first level, nodes are clustered into groups and connected via switched BSAs~\cite{koyamaOptimalSwitchingNetworks2024} to achieve all-to-all intra-group connectivity.
At the second level, nodes can route photons through their local group switch to a different group switch, enabling direct Bell pair generation between distinct groups.
While Sakuma et al.~\cite{sakumaQFlyOpticalInterconnect2025} considers a complete graph for the inter-group structure, the exact topology can vary and is best modeled as a regular chordal ring graph.
By starting with a standard ring and adding specific chordal shortcuts, the architecture minimizes the overall graph diameter.
Assuming this regular chordal structure, the complete Q-Fly architecture is uniquely defined by three parameters: the number of nodes per group, the total number of groups, and the specific set of inter-group chordal offsets.

\section{Architecture constraints}
\label{sec:arch-constraints}

In this work, we configure the Q-Fly architecture with 64 groups, where each group contains 12 computational nodes.
This specific configuration requires a maximum switch size of 18 ports, which is highly feasible for near-term photonic hardware and safely tolerates realistic coupling losses.
While the original Q-Fly proposal allows photons to traverse multiple switches, we strictly limit routing to a maximum of two switch hops to maintain high generation rates and fidelity.
We allocate approximately 1,000 physical qubits per node for computation.
This physical footprint is sufficient to host either a single logical surface code patch or a single code block of the bivariate bicycle (BB) code~\cite{bravyiHighthresholdLowoverheadFaulttolerant2024} at an appropriate distance, as current hardware projections indicate that individual devices can house thousands of physical qubits.
Under the BB code configuration, each node houses 10 logical qubits, allocating 9 for active computation and reserving 1 specifically for PBC extractors, following the model proposed by He et al.~\cite{heExtractorsQLDPCArchitectures2025}.

To minimize the inter-group network diameter and simplify routing logic, we define the inter-group chordal topology using a strict geometric progression of power-of-two offsets.
Specifically, group $j$ connects to groups $j+1$, $j+2$, $j+4$, $j+8$, $j+16$, and $j+32 \pmod{64}$.
Because the $+32$ offset is perfectly self-inverse modulo 64, it acts as a dedicated duplex channel that perfectly bisects the ring.
This specific chordal structure forms a directed circulant graph $C_{64}(1, 2, 4, 8, 16, 32)$.
Consequently, this topology allows any group to route to any other group in the network with a strict worst-case diameter of exactly 3 hops.
By bounding the network diameter to 3 hops, the architecture provides a good trade-off between low-latency teleportation and minimal optical switch overhead.

We further assume that each node possesses dedicated local magic state factories.
These states may be prepared using cultivation techniques coupled with necessary distillation rounds~\cite{gidneyMagicStateCultivation2024,litinskiMagicStateDistillation2019}.
We therefore conservatively assume that exactly one $\ket{T}$ state is available for consumption at each node during every logical code cycle.
We set the baseline code cycle time at 1$\mu s$.
Although decoding bivariate bicycle codes imposes stricter latency constraints than the surface codes, we identify the photonic interconnect speed as the primary system bottleneck.
Assuming a raw physical Bell pair generation rate of $10^5$ Hz and a $1/3$ distillation yield, the network sustains a logical entanglement rate of approximately $3.3 \times 10^4$ Hz~\cite{bonillaataidesConstantOverheadFaultTolerantBellPair2025,liHighRateHighFidelityModular2024}.
Because a single local logical operation requires $O(d)$ physical code cycles at a 1 $\mu s$ baseline, the local logical cycle time evaluates to exactly $d$ microseconds.
By comparing these best theoretical rates, the inter-node operations are $30/d$ times slower than that of local.
We can bound the network penalty to a factor of 2 to 10 for practical code distances.
These hardware specifications and operational rates form the rigid foundation for our subsequent resource estimation.

\section{Optimization Algorithms}
\label{sec:optimization-algs}

Solving optimization problems is a primary candidate for near-term quantum advantage in the early FTQC era, largely due to circuit depths that can scale linearly with the input size, contrasting with heavy cryptographic tasks like Shor's algorithm~\cite{shorPolynomialTimeAlgorithmsPrime1999}.
In this work, we focus on two specific algorithms: the quantum approximate optimization algorithm (QAOA)~\cite{farhiQuantumApproximateOptimization2014} and decoded quantum interferometry (DQI)~\cite{jordanOptimizationDecodedQuantum2025}.
We select these algorithms because recent numerical and theoretical evidence suggests they have the potential to achieve quantum advantage~\cite{goldenNumericalEvidenceExponential2023,boulebnaneSolvingBooleanSatisfiability2024,shaydulinEvidenceScalingAdvantage2024,khattarVerifiableQuantumAdvantage2025,rosmanisNearlyLineartimeDecoded2026}.
Additionally, they offer a diverse set of subroutines highly relevant for evaluation in the DQC setting.
Furthermore, they represent two distinct paradigms of quantum optimization: QAOA relies on Hamiltonian evolution to approximate the target ground state, whereas DQI utilizes amplitude encoding and the quantum Fourier transform reducing the optimization task to a classical decoding problem.

\subsection{Quantum Approximate Optimization Algorithm (QAOA)}
\label{sec:optimization-algs:qaoa}

QAOA is defined by two non-commuting Hamiltonians, a depth hyperparameter $p$, and sets of parameterized evolution angles $\vec{\gamma}$ and $\vec{\beta}$.
Specifically, the problem Hamiltonian ($H_P$) encodes the objective function to define the \emph{phase oracle}, while the mixing Hamiltonian ($H_M$) defines the \emph{mixer}.
The hyperparameter $p$ denotes the number of alternating iterations of these two unitary operations, and the angles govern their respective evolution times.
Together, these components prepare the final state
\begin{align}
    \ket{\psi(\vec{\beta}, \vec{\gamma})}_p= & \left(e^{-i \beta_p H_M} e^{-i \gamma_p H_P}\right)\left(e^{-i \beta_{p-1} H_M} e^{-i \gamma_{p-1} H_P}\right) \nonumber \\
& \times \cdots\left(e^{-i \beta_1 H_M} e^{-i \gamma_1 H_P}\right)\ket{\psi_0},
\end{align}
where $\ket{\psi_0}$ is the easily prepared ground state of the mixer.
Conceptually, implementing the phase oracle involves evaluating each clause of the objective function and rotating the phase of the computational basis states by a corresponding weight.
This evaluation can be executed sequentially, clause-by-clause, or in parallel given sufficient qubit routing and ancilla space.
Conversely, the mixer is most commonly implemented as a transverse-field Ising mixer, which simplifies down to parallel single-qubit rotations in the X basis.

Although QAOA is frequently misunderstood as a heuristic strictly for the noisy intermediate-scale quantum (NISQ) era, properties such as the scale-invariance of optimal angles and parameter transferability strongly suggest its scalability by eliminating the need to perform variational optimization for every instance~\cite{benchasattabuseAmplitudeAmplificationOptimization2022,sureshbabuParameterSettingQuantum2024}.
While rigorous theoretical work exploring the inner workings of QAOA remains sparse~\cite{bassoPerformanceLimitationsQAOA2022,farhiQuantumApproximateOptimization2022,hadfieldAnalyticalFrameworkQuantum2023,wurtzCounterdiabaticityQuantumApproximate2022,benchasattabuseLowerBoundsNumber2025}, strong empirical evidence suggests that QAOA provides a quantum advantage over classical solvers~\cite{goldenNumericalEvidenceExponential2023,shaydulinEvidenceScalingAdvantage2024}.
Furthermore, recent works demonstrate that QAOA applied to the Boolean satisfiability problem 8-SAT exhibits a threshold for quantum advantage over state-of-the-art classical solvers under specific assumptions~\cite{boulebnaneSolvingBooleanSatisfiability2024,omanakuttanThresholdFaulttolerantQuantum2025}.
For this reason, we evaluate the distributed circuit implementation for this specific SAT problem.
We follow the estimation in~\cite{omanakuttanThresholdFaulttolerantQuantum2025} and target problem instances near the algorithmic phase transition, characterized by a clause-to-variable ratio of $m/n \approx 176$, where satisfying assignments are classically hard to find~\cite{boulebnaneSolvingBooleanSatisfiability2024}.

To construct the phase oracle for an 8-SAT problem, the algorithm must evaluate the satisfiability of individual clauses, which typically take the form
\begin{equation}
    C_j = x_0 \lor x_1 \lor \cdots \lor x_7.
\end{equation}
The standard quantum implementation rewrites this boolean clause using De Morgan's laws as
\begin{equation}
    C_j = \lnot (\lnot x_0 \land \lnot x_1 \land \cdots \land \lnot x_7).
\end{equation}
This logical transformation allows the clause to be coherently evaluated by applying X gates to negate the relevant literals, followed by a multi-controlled Toffoli gate targeting an ancilla initialized in the $\ket{0}$ state.
Therefore, the primary subroutines required for our QAOA implementation are generalized boolean AND circuits to construct the phase oracle, and single-qubit rotations to construct the mixer.
While amplitude amplification can further boost success probabilities~\cite{omanakuttanThresholdFaulttolerantQuantum2025}, we omit this extension to maintain a foundational resource baseline for the distributed architecture.

\subsection{Decoded Quantum Interferometry (DQI)}
\label{sec:optimization-algs:dqi}

Unlike QAOA, which encodes the problem into the phase, DQI encodes the objective function into the amplitude and aims to prepare a state of the form (omitting the normalization factor)
\begin{equation}
    \ket{P(f)} = \sum_x P(f(x)) \ket{x},
\end{equation}
where $f(x)$ is the objective function and $P(f)$ is a carefully chosen polynomial.
The goal is to amplify the amplitudes of basis states corresponding to high objective values.
Because objective values can be negative, $P(f)$ must be designed to suppress small magnitudes and negative values while amplifying large positive ones.

Jordan et al.~\cite{jordanOptimizationDecodedQuantum2025} realized that for specific optimization problems like Max-LINSAT, the objective amplitudes become highly sparse when mapped to the Fourier domain.
Through Regev's reduction~\cite{regevLatticesLearningErrors2005}, preparing this sparse state is equivalent to decoding an underlying error-correcting code defined by this specific Max-LINSAT problem instance.
Although general decoding remains NP-hard, this formulation allows the algorithm to leverage established, efficient classical decoders---specifically LDPC codes for Max-LINSAT and Reed--Solomon codes for Optimal Polynomial Intersection (OPI) problems.
This framework has been generalized in~\cite{chaillouxQuantumDecodingProblem2024}, covering a broader family of codes and decoders.
Furthermore, recent work demonstrates that DQI can be implemented in near-linear time for OPI, hinting at verifiable quantum advantage~\cite{khattarVerifiableQuantumAdvantage2025,rosmanisNearlyLineartimeDecoded2026}.
While decoding can be approached via classical or quantum routes~\cite{chaillouxQuantumDecodingProblem2024}, utilizing a coherent implementation of a classical decoder offers additional insights: the algorithm's overall performance and solution quality (i.e., the approximation ratio) can be estimated simply by benchmarking the classical decoder itself.

For the circuit considered in this work, we adopt the implementation from~\cite{patamawisutQuantumCircuitDesign2025}, which provides a concrete circuit for DQI applied to Max-2-SAT (essentially a MaxCut problem) using Gauss-Jordan elimination for the decoding process.
While it has been shown that DQI provides no asymptotic advantage for MaxCut~\cite{parekhNoQuantumAdvantage2025}, it still serves as a good candidate circuit to study for DQC architecture.
The problem can be formally described as an instance of Max-LINSAT over $\mathbb{F}_2$, defined by a constraint matrix $B \in \mathbb{F}_2^{n \times m}$ and a target vector $v \in \mathbb{F}_2^m$.
Note that unlike standard formulations where a cut contributes $+1$ and no cut contributes $0$, for DQI objective functions, a satisfied clause (a cut) evaluates to $+1$, whereas an unsatisfied clause evaluates to $-1$.
The evolution of the states is given by (with the subscripts denoting the size of the registers)
\begin{align*} 
    \ket{0}
    &\xrightarrow{\text{Unary amplitude encoding}} \sum_{k=0}^{\ell} w_k \ket{k}_m \\
    &\xrightarrow{\text{Dicke state encoding}} \sum_{k=0}^{\ell} \hat{w}_k \sum_{\substack{y \in \mathbb{F}_2^m \\ |y|=k}} \ket{y}_m; \quad \left(\hat{w}_k = \frac{w_k}{\sqrt{\binom{m}{k}}}\right) \\
    &\xrightarrow{\text{Phase encoding}} \sum_{k=0}^{\ell} \hat{w}_k \sum_{\substack{y \in \mathbb{F}_2^m \\ |y|=k}} (-1)^{v \cdot y} \ket{y}_m \\
    &\xrightarrow{\text{Constraint encoding}} \sum_{k=0}^{\ell} \hat{w}_k \sum_{\substack{y \in \mathbb{F}_2^m \\ |y|=k}} (-1)^{v \cdot y} \ket{y}_m \ket{B^T y}_n \\
    &\xrightarrow{\text{Coherent decoding via Gauss-Jordan}} \sum_{k=0}^{\ell} \hat{w}_k \sum_{\substack{y \in \mathbb{F}_2^m \\ |y|=k}} (-1)^{v \cdot y} \ket{B^T y}_n \\
    &\xrightarrow{\text{Inverse QFT}} \sum_{x \in \mathbb{F}_2^n} P(f(x)) \ket{x}_n.
\end{align*}
Ideally, the decoding step (the second-to-last evolution) should uncompute the error register perfectly.
If the decoding is imperfect---meaning multiple corrupted codewords map to the same syndrome---the $\ket{y}_m$ register will not reliably return to $\ket{0}$ and remains entangled with the syndrome register, degrading the coherence of the final state.
This is why the coherent decoder acts as the performance bound for the algorithm.

To follow this evolution step-by-step: first, the weights $w_k$, derived from the chosen polynomial $P(f)$, are initialized via unary amplitude encoding.
These weights are then transformed into a superposed Dicke state, which represents all possible error configurations $\ket{y}$ with a fixed Hamming weight $k$.
After applying the target phase, the constraint matrix $B$ (whose dual is the parity-check matrix $B^T$ for this code) computes the parity into the syndrome register.
A coherent decoder then uncomputes the error register via Gauss-Jordan elimination.
Finally, an inverse quantum Fourier transform---which simplifies to a layer of Hadamard gates for binary fields---maps the state to the target amplitude encoding.

Therefore, the required subroutines for our DQI implementation include amplitude encoding preparation, Dicke state unitary, coherent LDPC decoding, and binary arithmetic (implemented via CNOTs).

\section{Optimization Techniques}
\label{sec:circuit-optimization-techniques}

Having established the conceptual foundations of QAOA and DQI, we now detail the specific circuit construction techniques and optimization strategies required for their efficient implementation in a distributed architecture.
Specifically, DQI requires the preparation of superposed Dicke states and coherent arithmetic, while 8-SAT QAOA implementation relies heavily on counting satisfied clauses and executing single-qubit rotations.
In this section, we examine these subroutines and demonstrate alternative space-time trade-offs within the DQC paradigm.

\subsection{Fan-out Operations}
\label{sec:fan-out-operations}

Fan-out, or a multi-target CNOT gate, is a powerful primitive used to construct and approximate many useful operations, such as the AND, parity, and OR gates~\cite{greenCountingFanoutComplexity2002,hoyerQuantumFanoutPowerful2005}.
In the Pauli-based computation (PBC) model, a fan-out operation costs the same as a single CNOT gate when implemented explicitly, meaning it can be executed cross-node using just one Bell pair per remote node.
Alternatively, if this operation is performed locally and absorbed into a non-Clifford multi-qubit Pauli measurement (MPP), it can be considered practically free.
These operations are useful for temporarily copying computational basis states across the architecture to parallelize commuting sequential controlled operations with the same controls.
As we will see in our specific application to the 8-SAT QAOA circuit, fan-out operations resolve scheduling bottlenecks when evaluating multiple clauses with overlapping qubit supports.
By fanning out the shared variables across the groups, we can evaluate these overlapping clauses more efficiently and in parallel.

\subsection{Addition}
\label{sec::quantum-addition}

Arithmetic circuits such as addition are core subroutines required by numerous other protocols, serving as essential building blocks for composing functions and calculations.
Let us first consider variants of quantum adder circuits.
The standard Gidney ripple-carry adder~\cite{gidneyHalvingCostQuantum2018} utilizes temporary logical-AND operations and measurement-based uncomputation (often called the TACU gadget) to minimize the non-Clifford cost, requiring $n - 1$ Toffoli gates and $n - 1$ ancillae for an $n$-bit addition.
While the Gidney adder is optimal in terms of its total $T$-gate count, our architecture restricts individual nodes to consuming exactly one $T$-state at a time.
Thus, even if we utilize the time-saving Toffoli construction from~\cite{jonesLowoverheadConstructionsFaulttolerant2013} where the $T$-depth is bounded to 1, executing a single Toffoli gate still requires 4 logical cycles.
Assuming these gates execute sequentially and that a fraction $r$ of the Toffoli gates require cross-node Bell pairs, we can estimate the total execution time,
\begin{equation}
    T_{\text{Gidney}} = (n - 1) (T_\text{Toff} + rT_{\text{Bell}}),
\end{equation}
where $T_{\text{Toff}}$ is the logical cycle time for implementing a Toffoli gate (4 in our case) and $T_{\text{Bell}}$ is the number of logical cycles per Bell pair consumption that we assume varies from 2 to 10, as discussed in \cref{sec:arch-constraints}.

An alternative adder with a higher total Toffoli count but a shallower Toffoli depth can decrease the overall execution time if the operations are parallelized across multiple nodes.
We can exploit this parallelization via the quantum carry-lookahead adder (QCLA) utilizing the Sklansky prefix tree structure~\cite{wangOptimalToffoliDepthQuantum2025,sklanskyConditionalSumAdditionLogic1960}.
This structure achieves a Toffoli depth of $\lceil \log_2 n \rceil + O(1)$ by fanning out intermediate carry bits to parallelize subsequent Toffoli operations.
After calculating the initial carry-propagate and carry-generate signals, the algorithm accumulates carries using a block-based fan-out structure that distributes the highest carry from a computed block simultaneously to all bits in the subsequent block.
This mechanism complements the Q-Fly architecture because strategically assigning contiguous logical blocks to different groups minimizes the required inter-group routing.
The chordal topology only needs to transmit a small number of carry bits between these distinct blocks per Toffoli layer, minimizing the consumption of cross-group Bell pairs.
Assuming the tree evaluates in $\lceil \log_2 n \rceil + 4$ layers and each layer requires parallel local Toffoli execution interspersed with Bell pair consumption, the execution time evaluates to
\begin{equation}
    T_{\text{QCLA}} = (\lceil \log_2 n \rceil + 4) (T_\text{Toff} + r T_{\text{Bell}}).
\end{equation}

These two addition strategies exhibit a clear space-time trade-off, allowing the architecture to flexibly prioritize either spatial efficiency via the Gidney adder or parallel execution speed via the QCLA depending on the immediate subroutine constraints and the routing overhead required to spatially align the operand blocks.

\subsection{Single-qubit rotations}
\label{sec:circuit-optimization-techniques:single-qubit-rotations}

The prevailing standard for single-qubit gate synthesis with the Clifford+T gate set is the \emph{gridsynth} algorithm~\cite{rossOptimalAncillafreeClifford+T2016}, an efficient realization of the Solovay-Kitaev theorem~\cite{dawsonSolovayKitaevAlgorithm2006}.
This method synthesizes arbitrary rotations by decomposing them into a sequence of Clifford ($H$, $S$, CNOT) and $T$-gates.
For rotations around the Z axis, the sequential execution time $T_{\text{Grid}}$ scales linearly with a target accuracy $\delta$ and required bit-precision $m = \lceil \log_2(1/\delta) \rceil$ as
\begin{equation}
    T_{\text{Grid}} = a + b m.
\end{equation}
Following the resource estimates in~\cite{omanakuttanThresholdFaulttolerantQuantum2025}, we adopt the parameter $(a, b) = (9.19, 3)$ based on the algorithms developed in~\cite{rossOptimalAncillafreeClifford+T2016}.
Because \emph{gridsynth} is inherently sequential, the availability of parallel magic states across the distributed architecture does not help accelerate its execution time.

An alternative approach leverages the phase kickback mechanism via the phase-gradient technique~\cite{kitaevClassicalQuantumComputation2002}.
This method relies on a pre-prepared, $m$-qubit phase gradient state $\ket{\phi_m}$, defined as
\begin{align}
    \ket{\phi_m} &= \frac{1}{\sqrt{2^m}} \sum_{k=0}^{2^m-1} e^{2\pi i k / 2^m} \ket{k} \nonumber \\
    &= \bigotimes_{j=1}^m \frac{1}{\sqrt{2}} \left( \ket{0} + e^{2\pi i / 2^j} \ket{1} \right).
\end{align}
This technique executes a relative phase rotation by an angle $\theta$ on a target state $\ket{\psi} = \alpha\ket{0} + \beta\ket{1}$ by temporarily computing the angle into an intermediate register conditioned on the target qubit, and subsequently adding it to the gradient state, as described by
\begin{align}
    &\ket{\psi}\ket{0}\ket{\phi_m} \nonumber \\
    &\quad \xrightarrow{\text{Controlled-Write } \theta} (\alpha\ket{0}\ket{0} + \beta\ket{1}\ket{\theta})\ket{\phi_m} \nonumber \\
    &\quad \xrightarrow{\text{Add to gradient}} \alpha\ket{0}\ket{0}\ket{\phi_m} + \beta\ket{1}\ket{\theta}\ket{\phi_m + \theta} \nonumber \\
    &\quad = \alpha\ket{0}\ket{0}\ket{\phi_m} + \beta e^{i\theta}\ket{1}\ket{\theta}\ket{\phi_m} \nonumber \\
    &\quad \xrightarrow{\text{Uncompute } \theta} (\alpha\ket{0} + \beta e^{i\theta}\ket{1})\ket{0}\ket{\phi_m}.
\label{eq:rotation-phase-gradient}
\end{align}
Because the addition operation acts cyclically on the gradient state, it kicks back the relative phase $e^{i\theta}$ onto the $\ket{1}$ component of the target state while returning the gradient state $\ket{\phi_m}$ to its original, disentangled form.
As a result, the gradient state is a reusable catalyst and can also be coherently copied across the architecture for further parallelization~\cite{gidneyActuallyYouCant}.

To efficiently execute this phasing via the phase-gradient method and addition, we utilize the QCLA.
We can determine the crossover point where this phasing via the phase-gradient method becomes advantageous over \emph{gridsynth} for a given target accuracy $\delta$ by solving the inequality $T_{\text{QCLA}} < T_{\text{Grid}}$, yielding 
\begin{equation}
    \left(\lceil \log_2(m) \rceil + 4 \right) (T_\text{Toff} + r T_{\text{Bell}}) < a + 3m.
\end{equation}
Assuming a conservative Bell pair consumption rate $T_{\text{Bell}} = 10$ and $r=1$, the crossover threshold occurs at just $m = 44$ bits of precision.
For implementation using 64 bits of precision, the QCLA-based phase gradient method completes in 140 cycles, saving 61 logical cycles per rotation compared to the 201 cycles required by \emph{gridsynth}.

\subsection{Linear phasing}
\label{sec:function-phasing}

Linear phasing is a direct extension of the phase gradient technique that embeds a constant scaling factor directly into the gradient state itself.
If we want to rotate the phase of a state by an angle proportional to the value in the register $\theta$ scaled by a constant $c$, we can first compute this explicit product using standard quantum arithmetic and apply the phase gradient method normally, expressed as
\begin{align}
    \ket{\theta} \ket{0} \ket{\phi_m} & \xrightarrow{\text{multiply}} \ket{\theta} \ket{c \cdot \theta} \ket{\phi_m} \nonumber \\
    & \xrightarrow{\text{add to gradient}} e^{2\pi i c \theta / 2^m} \ket{\theta} \ket{c \cdot \theta} \ket{\phi_m +  c \cdot \theta } \nonumber \\
    & \xrightarrow{\text{uncompute}} e^{2\pi i c \theta / 2^m} \ket{\theta} \ket{0} \ket{\phi_m}.
\end{align}
However, we can embed the weight directly into the gradient state because every angle is multiplied by the same constant.
By preparing a customized phase gradient state $\ket{\phi_m(c)}$ defined as
\begin{align}
    \ket{\phi_m(c)} &= \frac{1}{\sqrt{2^m}} \sum_{k=0}^{2^m-1} e^{2\pi i c k / 2^m} \ket{k} \nonumber \\
    &= \bigotimes_{j=1}^m \frac{1}{\sqrt{2}} \left( \ket{0} + e^{2\pi i c 2^{j-1} / 2^m} \ket{1} \right),
\end{align}
we achieve the exact same phase kickback by directly adding $\theta$ to this modified state.
Therefore, if we need to perform linear phasing multiple times, we can prepare this customized state ahead of time to save on multiplication circuits.

\subsection{Controlled rotations}
\label{sec:circuit-optimization-techniques:controlled-rotations}

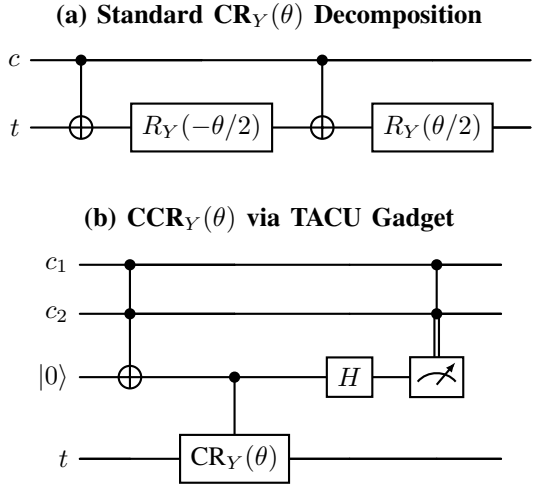
\begin{figure}[t]
    \centering
    
    \textbf{(a) Standard $\text{CR}_Y(\theta)$ Decomposition}\par\medskip
    \begin{quantikz}
        \lstick{$c$} & \ctrl{1} & \qw & \ctrl{1} & \qw & \qw \\
        \lstick{$t$} & \targ{} & \gate{R_Y(-\theta/2)} & \targ{} & \gate{R_Y(\theta/2)} & \qw
    \end{quantikz}
    
    \vspace{0.6cm}
    
    \textbf{(b) $\text{CCR}_Y(\theta)$ via TACU Gadget}\par\medskip
    \begin{quantikz}
        \lstick{$c_1$} & \ctrl{1} & \qw  &       &\ctrl{1} & \qw \\
        \lstick{$c_2$} & \ctrl{1} & \qw  &       &\control{} & \qw \\
        \lstick{$\ket{0}$}   & \targ{}  & \ctrl{1}&  \gate{H}  & \meter{} \wire[u][1]{c}   \\
        \lstick{$t$}   & \qw      & \gate{\text{CR}_Y(\theta)} & \qw & \qw & \qw
    \end{quantikz}
    
    \caption{Optimized synthesis of controlled continuous rotations. \textbf{(a)} The standard sequential fractional rotation. \textbf{(b)} The temporary-AND compute-uncompute (TACU) gadget, which intercepts multiple controls to bypass exponential angle fracturing. Measurement of the ancilla in the X-basis triggers a classical feed-forward $CZ$ correction on the controls.}
    \label{fig:controlled-rotations}
\end{figure}

Building upon the single-qubit rotation gates, we can construct controlled-rotation gates by composing CNOT gates and single-qubit rotations.
\Cref{fig:controlled-rotations}(a) shows the standard decomposition for a $\text{CR}_Y(\theta)$ gate requiring two rotations of $\theta/2$ on the target qubit.
Because these half rotations act sequentially on the identical target qubit, this standard circuit requires the temporal depth of two full rotation syntheses.

For multi-controlled operations such as $\text{CCR}_Y$ and $\text{CCR}_Z$ gates, we utilize the temporary-AND compute-uncompute (TACU) gadget to AND the control qubits into a clean ancilla, as shown in \cref{fig:controlled-rotations}(b).
We can then use a standard single-controlled $\text{CR}_Y$ gate, after which the ancilla is resolved via measurement-based uncomputation.
This approach bounds the non-Clifford overhead to a single Toffoli gate, and it can be generalized to an arbitrary number of controls by constructing a staircase of Toffoli gates to compute a global logical-AND prior to the rotation.

\subsection{Dicke State Unitary}
\label{sec:dicke-state-unitary}

Many previous works have investigated the preparation of Dicke states under various architectural scopes and constraints~\cite{bartschiDeterministicPreparationDicke2019,bartschiShortDepthCircuitsDicke2022,buhrmanStatePreparationShallow2024,liuLowdepthQuantumSymmetrization2025,grettaSuperConstantWeightDicke2026,yuEfficientPreparationDicke2026,piroliApproximatingManyBodyQuantum2024}.
While some approaches provide constant-depth circuits~\cite{buhrmanDistributedQuantumComputing2003,yuEfficientPreparationDicke2026,piroliApproximatingManyBodyQuantum2024}, they are not unitary, thus lacking the ability to directly generate a coherent superposition of Dicke states across differing weights, making them inapplicable for use in DQI.
While recent advances can prepare these superposed states~\cite{grettaSuperConstantWeightDicke2026,liuLowdepthQuantumSymmetrization2025} in very low depth, they demand large ancilla spaces (often $O(n \log n)$) for an $n$-bit Dicke state and complex arithmetic circuits via sorting, comparison, and threshold.
Furthermore, bounding the communication overhead and routing for these arithmetic circuits within a distributed architecture is highly complex, which we leave for future study.
To contain our analysis to a more easily manageable approach, we adopt the deterministic construction proposed by B\"{a}rtschi and Eidenbenz~\cite{bartschiShortDepthCircuitsDicke2022}, which defines a generalized Dicke state unitary $U_n$ building from a small size Dicke unitary proposed in~\cite{bartschiDeterministicPreparationDicke2019}.
When provided with an $n$-qubit input register prepared in a unary basis of weight $k$, this unitary distributes the amplitude to generate the corresponding Dicke state, defined by
\begin{equation}
    U_n \ket{1^k 0^{n-k}} = \ket{D_k^n} = \binom{n}{k}^{-1/2} \sum_{x \in \{0,1\}^n, |x|=k} \ket{x}.
\end{equation}

The construction uses a divide-and-conquer strategy to distribute the $1$s bits from a single primary register into recursively smaller sub-registers.
This weight distribution process is realized by the \emph{Weight Distribution Block} (WDB) unitary.
This recursion creates a globally entangled state where each sub-register contains the correct number of $1$s bits with the appropriate weight distribution.
Dicke unitaries are then applied to these sub-registers transforming them into the final Dicke states.
For the two-register case, defining the primary register size as $l = n-m$, this sequence can be shown by
\begin{align}
    &\ket{1^k 0^{l-k}}_{l}\ket{0}_m \nonumber \\
    &\quad \xrightarrow{\text{WDB}_{k}^{n,m}} \sum_{j=0}^{\min(k,m)} w_{k,j} \ket{1^{k-j} 0^{l-(k-j)}}_{l} \ket{1^j 0^{m-j}}_m \nonumber \\
    &\quad \xrightarrow{U_{l} \otimes U_m} \sum_{j=0}^{\min(k,m)} w_{k,j} \ket{D_{k-j}^{l}} \otimes \ket{D_j^m} = \ket{D_k^{n}},
\label{eq:short-depth-dicke}
\end{align}
where $w_{k,j} = \sqrt{\binom{m}{j}\binom{l}{k-j} / \binom{n}{k}}$, the subscripts denote the register sizes, and the $\text{WDB}_{k}^{n,m}$ operator distributes up to $k$ $1$s bits from the total $n$-qubit space into the $m$-qubit target split.
When distributing to more than two registers, the WDB blocks are applied recursively.
This recursive splitting process is repeated until reaching a predefined stopping point, such as a 3-qubit block, at which point a Dicke unitary $U_3$ is applied to all sub-registers to generate the full Dicke state.

\begin{figure}[t]
    \centering
    \begin{quantikz}[row sep=0.4cm, column sep=0.6cm]
        \lstick{$ctrl$} & \ctrl{1}                  & \ctrl{2}      & \ctrl{5}      & \qw \\
        \lstick{$u_1$}  & \gate{R_Y(\theta_{n,1})}  & \ctrl{1} & \qw      & \qw \\
        \lstick{$u_2$}  & \qw                       & \gate{R_Y(\theta_{n,1})} &  & \qw \\
        \wave{}&&&&& \\
        \lstick{$u_{n-1}$}  & \qw                       & & \ctrl{1}      & \qw \\
        \lstick{$u_n$}  & \qw                       & & \gate{R_Y(\theta_{n,n})}      & \qw
    \end{quantikz}
    \caption{Partial circuit illustrating the controlled amplitude distribution ladder for the $n$-th qubit.}
    \label{fig:dicke-amplitude-ladder}
\end{figure}
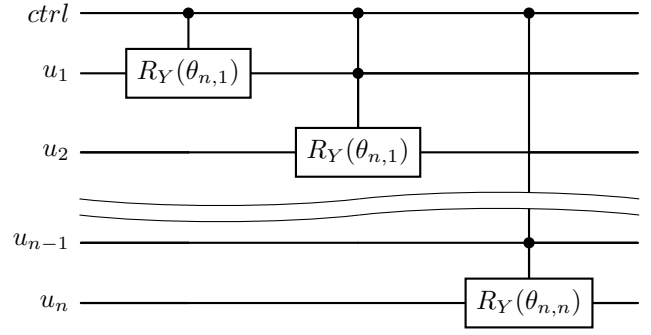

To conceptualize the workings of a WDB, let $l = n-m$ denote the size of the primary sub-register, and let $\ket{k}_l$ denote its unary encoding of weight $k < l$.
When a WDB distributes up to $k$ 1s bits into an $m$-qubit secondary sub-register, the state evolution proceeds through three stages.
Given an arbitrary input superposition encoded in unary as $\sum_k \alpha_k \ket{k}_l$, the circuit first initializes a clean $m$-qubit secondary register.
The primary register then changes from a unary to a one-hot encoding by retaining only the most significant bit of the $1$s string.
Utilizing a staircase pattern of $\text{CCR}_Y$ gates controlled by the $k$-th one-hot qubit, as illustrated in \cref{fig:dicke-amplitude-ladder}, the block reads the primary register and writes a superposition of partitioned weights $j$ into the secondary register.
We then revert the one-hot back to unary.
The full distribution can be described by
\begin{equation}
    \sum_k \alpha_k \ket{k}_l \ket{0}_m \xrightarrow{\text{distribute}} \sum_k \alpha_k \sum_{j=0}^{\min(k,m)} w_{k,j} \ket{k}_l \ket{j}_m,
\end{equation}
where both $j$ and $k$ are represented in unary (omitting the intermediate one-hot conversion steps for clarity), and the amplitude weights $w_{k,j}$ follow the previously defined hypergeometric distribution.
A subsequent unary subtraction circuit then removes this distributed weight $j$ from the primary register, yielding the final state
\begin{equation}
    \xrightarrow{\text{subtract}} \sum_k \alpha_k \sum_{j=0}^{\min(k,m)} w_{k,j} \ket{k-j}_{l} \ket{j}_m,
\end{equation}
completing the exact state partition detailed in the second step of the transformation in \cref{eq:short-depth-dicke}.

To quantify the resource overhead, suppose that all arbitrary rotations are decomposed using \emph{gridsynth}.
We assume that operations within a single WDB block evaluate sequentially, but that independent WDB blocks at the same level of recursion execute in parallel.
During the distribution phase, a single WDB block separating a maximum weight $k$ evaluates a triangular ladder of $\text{CCR}_Y$ gates, requiring $k + (k-1) + \dots + 1 = \frac{k(k+1)}{2}$ distinct controlled rotations.
Because generating an $n$-qubit Dicke state with a maximum weight $k$ requires approximately $\lceil \log_2 k \rceil$ layers of parallel WDB operations, the total sequential $\text{CCR}_Y$ depth evaluates to $\frac{1}{2} k(k+1) \lceil \log_2 k \rceil$.
Furthermore, the subsequent unary subtraction circuits require the exact same depth of Toffoli gates.
Therefore, we can bound the execution time by scaling this operation depth by the time required to execute a \emph{gridsynth} rotation and the necessary Toffoli gates, incorporating the inter-node routing penalty $r T_{\text{Bell}}$, yielding
\begin{equation}
    T_{\text{Dicke}} \approx \frac{1}{2} k(k+1) \lceil \log_2 k \rceil \Big( T_{\text{Grid}} + 2(T_\text{Toff} + rT_{\text{Bell}}) \Big).
\end{equation}

\section{Resource Estimation}
\label{sec:resource-estimation}

\begin{table*}[t]
    \centering
    \caption{Resource estimation of core subroutines and algorithms.
    The variable $r$ in the analytical formulas denotes the ratio of Bell pairs consumed per Toffoli gate to perform cross-node Toffoli operations, fan-outs, or teleportations.
    Because this ratio varies based on the subroutine's specific qubit placement and communication density, temporal costs are evaluated utilizing subroutine-specific average-case routing ratios ($r=1$ for the QCLA; $r=1/3$ for the rest).
    QAOA and DQI stages are evaluated using their respective algorithmic parameters ($n=64, m=11,264$ for QAOA; $n=50, m=200$ for DQI).
    The rotations in both QAOA and DQI are executed with 64-bit precision.
    Columns labeled AV$_2$ and AV$_{10}$ represent the theoretical lower bounds for the Active Volume architecture~\cite{litinskiActiveVolumeArchitecture2022}, which exclude reaction time and account only for the number of blocks processed within our architecture evaluated at $T_{\text{Bell}} = 2$ and $T_{\text{Bell}} = 10$, respectively.}
    \label{tab:resource-estimation}
    \renewcommand{\arraystretch}{1.3}
    \begin{tabulary}{\textwidth}{p{0.17\textwidth} p{0.34\textwidth} c c c c c}
        \toprule
        \multirow{2}{*}{\begin{tabular}{@{}l@{}}\textbf{Subroutine /} \\ \textbf{Algorithm Stages}\end{tabular}} & \multirow{2}{*}{\begin{tabular}{@{}l@{}}\textbf{Analytical Formula} \\ \textbf{(Logical Cycles)}\end{tabular}} & \multicolumn{5}{c}{\textbf{Temporal Execution Time (Logical Cycles)}} \\
        \cmidrule(lr){3-7}
        & & \textbf{AV}$_2$ & $T_{\text{Bell}} = 2$ & $T_{\text{Bell}} = 5$ & \textbf{AV}$_{10}$ & $T_{\text{Bell}} = 10$ \\
        \midrule
        
        \multicolumn{7}{l}{\textbf{Core Subroutines}} \\
        Gidney Adder ($n$-bit) & $(n - 1) \big( T_{\text{Toff}} + r T_{\text{Bell}} \big)$ & 19 & 294 & 357 & 94 & 462 \\
        QCLA (Sklansky, 64-bit) & $(\lceil \log_2 n \rceil + 4) (T_{\text{Toff}} + r T_{\text{Bell}})$ & --- & 60 & 90 & --- & 140 \\
        Gridsynth Rotation (64-bit) & $\approx 201$ & 32 & 201 & 201 & 156 & 201 \\
        Dicke State Unitary ($k=25$, 64-bit phasing) & $\approx \frac{1}{2} k(k+1) \lceil \log_2 k \rceil \big( T_{\text{Grid}} + 2(T_\text{Toff} + rT_{\text{Bell}}) \big)$ & --- & 341,792 & 345,042 & --- & 350,458 \\
        
        \midrule
        \multicolumn{7}{l}{\textbf{QAOA Iteration Stages (8-SAT, $n=64$, $m=11,264$)}} \\
        Intra/Inter-Group Fan-out & $79 T_{\text{Bell}}$ & --- & 158 & 395 & --- & 790 \\
        Clause Evaluation & $176 \big(\max(7 T_{\text{Bell}}, 5 T_{\text{Toff}}) + T_{\text{Grid}}\big)$ & 396,294 & 38,896 & 41,536 & 1,936,410 & 47,696 \\
        Mixer Rotations & $T_{\text{Grid}}$ & 2,048 & 201 & 201 & 9,984 & 201 \\
        \rowcolor{gray!10} \textit{Total QAOA Iteration} & \textit{Sum of QAOA stages} & 398,342 & 39,255 & 42,132 & 1,946,394 & 48,687 \\

        \midrule
        \multicolumn{7}{l}{\textbf{DQI Execution Stages ($n=50, m=200$, $l = 25$)}} \\
        Setup \& Unary Encoding & $201 + 24(2 T_{\text{Bell}} + T_{\text{QCLA}})$ & 1,633 & 1,737 & 2,601 & 7,962 & 4,041 \\
        Dicke Preparation ($l=25$) & $1625 \big(209 + \frac{2}{3}T_{\text{Bell}}\big)$ & 873,077 & 341,792 & 345,042 & 4,256,369 & 350,458 \\
        Constraint Encoding & $2m T_{\text{Bell}} = 400 T_{\text{Bell}}$ & 208 & 800 & 2,000 & 1,042 & 4,000 \\
        Syndrome Decoding & $2550 T_{\text{Bell}}$ & 208 & 5,100 & 12,750 & 1,042 & 25,500 \\
        \rowcolor{gray!10} \textit{Total DQI Execution} & \textit{Sum of DQI stages} & 875,126 & 349,429 & 362,393 & 4,266,415 & 383,999 \\
        \bottomrule
    \end{tabulary}
\end{table*}

Now that we have established the subroutines required to compose QAOA and DQI, we can estimate their resource overheads and total runtime.
To accurately evaluate the temporal cost of these algorithms within a distributed Q-Fly architecture, we must account for the qubit placement and the network routing overhead of rearranging qubits alongside the base subroutine costs.
Our previous subroutine estimates assumed an optimal placement of logical qubits—particularly for the QCLA, where the corresponding bits of the two addends must reside within the same node to properly utilize the space and propagate the carries.
To model this, we divide each algorithmic stage into a spatial rearrangement phase followed by the subroutine execution.
Hence, the temporal cost of any stage evaluates to the sum of its rearrangement routing delay and its subroutine execution time.
We provide the comprehensive analytical scaling and the final numerical cycle costs for all subroutines and algorithmic stages in \cref{tab:resource-estimation}.

\subsection{Estimation with 8-SAT QAOA}
\label{sec:estimation-with-8-sat-qaoa}

Assuming that the 8-SAT instances are randomly generated at the satisfiability threshold, we consider $n = 64$ variables, which contains $m = 64 \times 176 = 11,264$ clauses.
We also assume that the rotation angles are fixed and provided by the framework in~\cite{boulebnaneSolvingBooleanSatisfiability2024}.

A single iteration of the QAOA circuit consists of alternating applications of a phase oracle $U_P(\gamma)$ and a mixer $U_M(\beta)$.
To maximize our architectural space usage, we initially distribute the SAT variables such that variable $x_j$ is uniquely assigned to node group $j$.
To execute the phase oracle, we proceed through the following stages.

\paragraph{Intra-group fan-out} We first fan out variable $x_j$ locally within its home group $j$ to create a localized GHZ state, leveraging the non-blocking intra-group switches to complete this operation in exactly 2 $T_{\text{Bell}}$.

\paragraph{Inter-group all-to-all broadcast} We subsequently distribute these variables across the entire network so that every group possesses a full copy of all 64 variables.
Each group can utilize its chordal links to fan out variables to exactly 6 distinct groups per logical cycle, requiring $\lceil 63 / 6 \rceil = 11$ sequential $T_{\text{Bell}}$ routing steps.
Afterward, we use measurement-based uncomputation to remove duplicate variables within the same group.
This leaves exactly 64 distinct variables per group, which we distribute across the available computational nodes such that each node holds approximately 7 variables alongside at least 2 free ancilla qubits.
This final rearrangement incurs a time cost of up to 64 $T_{\text{Bell}}$, bringing the total temporal cost of this broadcast step to 75 $T_{\text{Bell}}$.

\paragraph{Clause evaluation and direct phasing} With the variables arranged, the nodes locally evaluate the 8-SAT clauses using multi-controlled Toffoli (MCT) gates.
On average, evaluating the partial clause components within a single node requires approximately 5 Toffoli gates~\cite{khattarRiseConditionallyClean2025}.
During this local MCT execution, the Bell pairs can be generated in the background.
We then fan out and combine these partial evaluations into a single variable, imposing a total cost bounded by $\max(7 T_{\text{Bell}}, 5 T_{\text{Toff}})$ to account for gathering results from up to 7 nodes.
For each satisfied clause, we perform the phasing via \emph{gridsynth} with angle $\gamma$ specified for the current QAOA iteration with $T_\text{Grid} = 201$ (\cref{sec:circuit-optimization-techniques:single-qubit-rotations}).
Following the phasing, the cross-node temporary-ANDs are uncomputed via measurement without any time penalty, while the internal node clean-up requires an average of 5 Toffoli gates.
Because the we can prepare Bell pairs for the subsequent round during the phasing, the pipeline effectively hides the communication latency.
The total time for this step evaluates to an initial 2 $T_{\text{Bell}}$ start-up cost plus the pipelined execution of $176 \times (\max(7 T_{\text{Bell}}, 5 T_{\text{Toff}}) + T_\text{grid})$.

\paragraph{Uncomputation and mixer} Following the phase oracle, we uncompute the variable fan-outs using measurement-based uncomputation, which instantly clears the network without incurring any additional temporal cost.
For the final mixer application $U_M(\beta)$, applying rotations across all 64 variables via the distributed phase gradient state would induce a massive inter-group routing bottleneck.
Instead, we synthesize these mixer rotations locally using \emph{gridsynth}, requiring a constant depth of $T_{\text{Grid}} = 201$ logical cycles.
Combining these stages, the total time cost to execute a single iteration of the QAOA circuit evaluates to
\begin{align}
    T_{\text{total}} &= 79 T_{\text{Bell}} \nonumber + 176 \max(7 T_{\text{Bell}}, 5 T_{\text{Toff}}) \\
    &\quad + 176\max(156 T_{\text{Toff}}, 19 T_{\text{Bell}})+ T_{\text{Grid}},
\end{align}
where $T_{\text{Grid}} = 201$ logical cycles.

\subsection{Estimation with DQI}
\label{sec:estimation-with-dqi}

Unlike QAOA, which evaluates its phase oracle in a highly parallelized manner, the execution of DQI is inherently sequential as detailed in \cref{sec:optimization-algs:dqi}.
To estimate the resources for a Max-LINSAT problem in the binary field, we consider a problem instance with $n = 50$ variables and $m = 200$ clauses, executing all rotations with 64 bits of accuracy.
To construct the complete DQI circuit, we proceed through the following stages as established in~\cite{patamawisutQuantumCircuitDesign2025}.

\paragraph{Setup stage} We first prepare a 64-bit standard phase gradient state by initializing one dedicated qubit within each of the 64 node groups.
As previously detailed in \cref{sec:circuit-optimization-techniques:single-qubit-rotations}, this preparation requires an execution time of 201 logical cycles.

\paragraph{Unary state amplitude encoding} Following the preparation of the gradient state, we prepare the amplitude encoding of the unary state, which is a sequential staircase comprising an initial $R_y$ rotation followed by $l-1$ $CR_y$ gates (for weight $l$).
Because DQI does not use the massive initial variable fan-out like QAOA, we retain abundant ancilla space to host the 64-bit phase gradient state locally across the network for utilization with phasing via QCLA (\cref{sec:circuit-optimization-techniques:controlled-rotations}).
We execute each $CR_y$ rotation by first writing the rotation angle $\theta$ to the angle register, fanning out the control qubit to every group via a global GHZ state that costs 2 $T_{\text{Bell}}$.
Then we apply the phase kickback using the QCLA adder, which requires $T_{\text{QCLA}}$ logical cycles.
Because the algorithm uncomputes the angle $\theta$ via measurement-based uncomputation without any time penalty, the total cost evaluates to $T_{\text{Unary}} = (l - 1)(2 T_{\text{Bell}} + T_{\text{QCLA}})$.

\paragraph{Dicke state preparation} We now need to turn the unary states into superposed Dicke states $\ket{D_m^l}$ over the $m = 200$ clause qubits.
While the Hamming weight $l$ satisfies $l < m/2$, implementations utilizing optimized polynomials $P(f)$ to maximize the approximation ratio typically require significantly smaller weights~\cite{jordanOptimizationDecodedQuantum2025,khattarVerifiableQuantumAdvantage2025}.
Assuming a conservative worst-case weight of $l = 25$, its time cost (given in \cref{sec:dicke-state-unitary}) evaluates to
\begin{equation}
    T_{\text{Dicke}} \approx \frac{1}{2} l(l+1) \lceil \log_2 l \rceil \Big( T_{\text{Grid}} + 2(T_{\text{Toff}} + rT_{\text{Bell}}) \Big).
\end{equation}

\paragraph{Constraint encoding and phasing} Next, we encode the problem constraints by applying the transposed parity matrix $B^T$, which translates corrupted codewords ($m$-register Dicke state) to syndrome ($n$-register).
Assuming the syndrome register resides in entirely distinct nodes from the Dicke state register, this non-local mapping requires up to $2m$ cross-node Bell pairs.
By utilizing GHZ fan-out techniques, the time cost scales proportionally to the number of clauses, given by $2m T_{\text{Bell}}$.

\paragraph{Syndrome decoding} To decode the errors via Gauss-Jordan Elimination (GJE), we need to apply up to $mn$ CNOTs alongside a maximum of $n$ SWAPs.
To transform the syndrome back into the corrupted codeword, we first compute the row echelon form as detailed in~\cite[Section.~VI]{patamawisutQuantumCircuitDesign2025}.
Recall that GJE proceeds by first finding the pivot and using that pivot to do row operations to eliminate all 1s below it, column-by-column.
This pivoting requires up to $n$ SWAP operations within the syndrome register, imposing a worst-case routing cost of $2n T_{\text{Bell}}$.
The subsequent row elimination requires up to $m-1$ CNOT operations per variable, but because we pack the error register contiguously across nodes, this requires at most $\lceil m / 9 \rceil T_{\text{Bell}}$ per column, evaluating to $n \lceil m / 9 \rceil T_{\text{Bell}}$ over all $n$ iterations.
To subsequently uncompute the corrupted codeword, we consume up to $n$ cross-node CNOT gates, costing an additional $n T_{\text{Bell}}$.
Finally, reversing this entire decoding computation doubles the row echelon execution overhead, bounding the total time cost to $T_{\text{Decode}} \le \big( 2(2n + n \lceil m/9 \rceil) + n \big) T_{\text{Bell}}$.

\paragraph{Hadamard transform and measurement} In the final step of the algorithm, we apply the inverse Quantum Fourier Transform (iQFT) over the binary field.
Because this binary iQFT simplifies to a layer of Hadamard gates on all $n$ qubits, it incurs zero logical cycles.

Summing these stages reveals that the total runtime of DQI is overwhelmingly dominated by the arithmetic depth of the Dicke state preparation.
Combining the components derived above, the total temporal cost to execute the entire DQI algorithm bounds to
\begin{align}
    T_{\text{total}} &\le 201 + (n - 1)(2 T_{\text{Bell}} + T_{\text{QCLA}}) + T_{\text{Dicke}} \nonumber \\
    &\quad + \bigg( 2m + 5n + 2n \left\lceil \frac{m}{9} \right\rceil \bigg) T_{\text{Bell}}.
\end{align}

\subsection{Comparison of execution time with surface code architecture}
\label{sec:compare-with-surface-code-active-volume}

To establish a comparative baseline, we assume the theoretical lower bound of a surface code compiled within an active volume architecture~\cite{litinskiActiveVolumeArchitecture2022}, which represents the current state of the art for execution time.
Strictly speaking, our Q-Fly architecture lacks the quick-swap operations required to natively support an active volume compilation, and its restricted topology would limit the flexible, high-throughput block routing assumed in~\cite{litinskiActiveVolumeArchitecture2022}.
Nevertheless, we establish a highly generous lower bound for the surface code by directly comparing the raw physical qubit counts and completely ignoring the non-negligible routing and reaction times.
Because our architecture provides exactly $12 \times 64$ logical qubits under the BB code, and the active volume model divides this equivalent spatial footprint equally between workspace and compute regions, the surface code can theoretically consume a maximum of $6 \times 64 = 384$ logical blocks per cycle.
Because cross-node operations require remote lattice surgery, we assume that a single logical cycle in the active volume takes $1 T_{\text{Bell}}$ instead of the 1 logical cycle unit used in our architecture.
We use this theoretical maximum throughput to calculate the active volume baseline values populated in \cref{tab:resource-estimation} normalized to our definition of logical cycles.

At first glance, the execution times for the individual active volume subroutines appear exceptionally low because these operations are highly parallelized across the available spatial blocks.
However, when composing these subroutines into complete algorithms like QAOA or DQI, the active volume architecture must process the total gate count rather than the circuit depth.
To ensure a fair comparison, we minimize the active volume's gate count by using the Dicke state construction from~\cite{bartschiDeterministicPreparationDicke2019} rather than our architecture's short-depth version~\cite{bartschiShortDepthCircuitsDicke2022}, and by fully parallelizing the QAOA clause evaluation with Gidney adders.
We adopt the overall gate counts from~\cite[Table~I]{patamawisutQuantumCircuitDesign2025}.
Additionally, we extract each gate block costs directly from~\cite[Table~1]{litinskiActiveVolumeArchitecture2022}.

As shown in \cref{tab:resource-estimation}, although individual active volume subroutines initially seem cheap, our BB-coded architecture outperforms the surface code by over an order of magnitude for both complete algorithms.
For instance, under $T_{\text{Bell}} = 10$, a single QAOA iteration demands nearly two million cycles in the active volume architecture, whereas our Q-Fly architecture completes the same iteration in fewer than 50,000 logical cycles.
This massive performance gap perfectly highlights the fundamental space-time tradeoff.
By shifting to a more space-efficient error correction code on the exact same physical hardware, we tap into substantially more logical capacity to perform operations in parallel, accelerating the overall execution time.
If the surface code were granted ten times the physical hardware to match the logical qubit capacity of our BB code, its execution time would decrease by a factor of 10.
Under this scaled scenario, the surface code would remain slower than our BB code when the network routing is slow ($T_{\text{Bell}} = 10$), but it would become comparable, if not faster, when the routing is fast ($T_{\text{Bell}} = 2$).
Finally, we must acknowledge that while our proposed circuit constructions inherently favor depth-optimized architectures, completely distinct circuit constructions tailored for active volume compilation likely exist and could further optimize the surface code baseline.

\section{Conclusion}
\label{sec:discussion}

In this work, we demonstrated that operating Pauli-based computation over high-rate qLDPC codes offers a substantial time advantage in distributed quantum architectures, outperforming the state-of-the-art active volume surface code model by over an order of magnitude given identical physical qubit counts even with generous estimations.
By exploiting the abundant logical space provided by the high-rate code, we designed parallelized subroutines for QAOA and DQI that bypass the sequential bottlenecks typically associated with monolithic PBC.
Contrary to recent proposals that optimize monolithic systems by minimizing overall qubit counts~\cite{websterPinnacleArchitectureReducing2026,mundadaHeterogeneousArchitecturesEnable2026,cainShorsAlgorithmPossible2026}, our approach treats total qubit count as an abundant computational resource.
Because distributed quantum computing is primarily limited by individual node capacities and interconnect speeds, deliberately trading this space for temporal acceleration becomes an essential design principle.

To support this abundance of qubits without inducing communication bottlenecks, the architecture requires an efficient network topology.
Deep multi-tiered designs like BCube, fat-trees, or Clos networks~\cite{shapourianQuantumDataCenter2025,pouryousefBenchmarkingQuantumData2026} demand massive switch sizes and strictly bounded optical losses to overcome the coupling penalties of traversing multiple layers.
We demonstrated that a two-level hierarchical structure like the Q-Fly architecture provides sufficient connectivity to guarantee high routing speeds without imposing these severe hardware constraints.
Crucially, unlocking the full potential of such distributed topologies requires a paradigm shift in circuit design.
Rather than merely partitioning circuits originally optimized to conserve space, we must focus on constructing novel, architecture-aware circuits.
Our work establishes that actively leveraging the space-time tradeoff via high-rate codes is a highly practical and essential strategy for realizing the full advantage of distributed quantum computation.

\bibliographystyle{IEEEtran.bst}
\bibliography{IEEEabrv, bibfile}

\end{document}